\documentclass[prl,twocolumn,superscriptaddress,floatfix,showpacs]{revtex4}

\usepackage{psfig}

\begin{document}
\title{Dynamical Properties of Charged Stripes in La$_{2-x}$Sr$_x$CuO$_4$}
\author{L. Tassini}
\affiliation{Walther Meissner Institut, Bayerische Akademie der
Wissenschaften, 85748 Garching, Germany}
\author{F. Venturini}
\altaffiliation[Permanent address: ]{Bruker Biospin AG, 8117
F\"allanden, Switzerland} \affiliation{Walther Meissner Institut,
Bayerische Akademie der Wissenschaften, 85748 Garching, Germany}
\author{Q.-M. Zhang}
\altaffiliation[Permanent
address: ]{National Laboratory of Solid State Microstructures,
Department of Physics, Nanjing University, Nanjing 210093, P. R.
China}
\affiliation{Walther Meissner Institut, Bayerische Akademie
der Wissenschaften, 85748 Garching, Germany}
\author{R. Hackl}
\affiliation{Walther Meissner Institut, Bayerische Akademie der
Wissenschaften, 85748 Garching, Germany}
\author{N. Kikugawa}
\altaffiliation[Permanent address: ]{Department of Physics, Kyoto
University, Kyoto 606-8502, Japan}
\affiliation{ADSM, Hiroshima
University, Higashi-Hiroshima 739-8526, Japan}
\author{T. Fujita}
\altaffiliation[Also at: ]{Institute of Spatial Science for
Regional and Global Culture, Waseda University, Tokyo 169-8555,
Japan}
\affiliation{ADSM, Hiroshima University, Higashi-Hiroshima
739-8526, Japan}
\date{\today}

\begin{abstract}
Inelastic light-scattering spectra of underdoped ${\rm
La_{2-x}Sr_xCuO_4}$ (LSCO) single crystals are presented which
provide direct evidence of the formation of quasi one-dimensional
charged structures in the two-dimensional $\rm CuO_2$ planes. The
``stripes'' manifest themselves in a Drude-like peak at low
energies and temperatures. The selection rules allow us to
determine the orientation to be along the diagonals at $x=0.02$
and along the principal axes at $x=0.10$. The electron-lattice
interaction determines the correlation length which turns out to
be larger in compound classes with lower superconducting
transition temperatures. Temperature is the only scale of the
response at different doping levels demonstrating the importance
of quantum critical behavior.

\end{abstract}
\pacs{74.72.-h, 74.20.Mn, 78.30.-j, 78.67.-n}

\maketitle

\begin{figure}[b]
\centerline{\psfig{figure=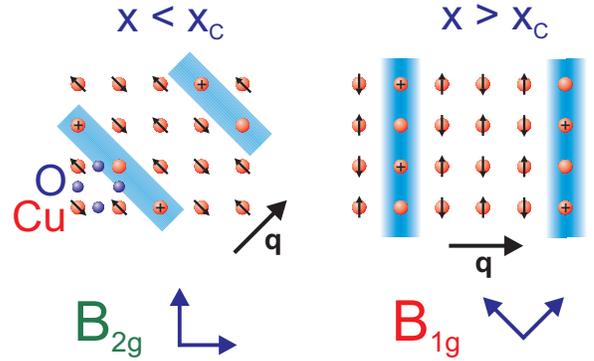}}

\caption[]{Sketch of spin-charge-ordered states in the
copper-oxygen plane (adopted from ref.~\cite{tra95}). There are
antiferromagnetic insulating areas and charged spin-free
``stripes''. The modulation is characterized by the vector {\bf
q}. If the pattern fluctuates the correlation length $\xi_s$ is
finite and can be as small as a few lattice constants. The
response of 1D objects perpendicular to the modulation direction
{\bf q} can only be observed by Raman scattering if the
polarization vectors of both the incoming and the outgoing photons
have a finite projection on either {\bf q} or the stripe direction
for transverse or longitudinal excitations, respectively. This
implies that stripes parallel to the principal axes can be
observed only in $B_{1g}$ and diagonal ones only in $B_{2g}$
symmetry. In either orientation orthogonal stipes are equivalent
and cannot be distinguished.}

\label{stripes}
\end{figure}

Spin-charge separation is a well-known phenomenon in
one-dimensional (1D) conductors \cite{lut63}. It has also been
proposed to occur in the essentially two-dimensional (2D)
copper-oxygen planes of high-temperature superconductors
\cite{zaa89, machida90, and91}. Under certain circumstances static
charged ``stripes'' in an antiferromagnetically ordered matrix
\cite{tra95} are observed such as sketched in Fig.~\ref{stripes}.
The phenomenon can be envisaged as a periodic charge modulation
best comparable to a density wave. Moreover, there are several
scenarios in which charge and spin ordering play a pivotal role in
explaining superconductivity \cite{cas97,car02}. In particular,
charge ordering fluctuations would be capable of providing an
effective mechanism for the formation of Cooper pairs
\cite{cas97}. Therefore, apart from being an interesting
phenomenon in itself the understanding of the dynamics of stripes
is an important issue in the cuprates.

In recent years a great variety of methods has been employed to
study stripes \cite{kiv03}. In underdoped cuprates regular
patterns in the electron distribution can eventually be observed
in the pseudogap state \cite{sta99} by scanning tunneling
microscopy \cite{ver04}. Infrared spectroscopy (IRS) and inelastic
light scattering show the {\em dynamical} behavior particularly
well. Strong peaks at low but finite energy develop in the
conductivity $\sigma^{\prime}(\omega,T)$ and in the Raman response
$\chi^{\prime \prime}(\omega,T)$ \cite{dum02,ven02,luc03}.
Explicit calculations demonstrate the existence of an absorption
at low energy due to a transverse excitation of charged stripes
\cite{ben03}. As in the case of IRS the Raman spectra and their
temperature dependence observed in ${\rm La_{1.90}Sr_{0.10}CuO_4}$
\cite{ven02} are similar to the results in the ladder compound
${\rm Sr_{14}Cu_{24}O_{41}}$ \cite{osa99,goz03}. Therefore, it is
qualitatively clear which type of response one can expect in
systems with a tendency to form charged stripes.

The interpretation of the Raman spectra of ${\rm
La_{1.90}Sr_{0.10}CuO_{4}}$ in terms of fluctuating stripes
\cite{ven02} can be tested via the selection rules. For stripe
orientations along the principal axes and the diagonals of the
copper-oxygen planes the response is expected in the $B_{1g}$
$(x^2-y^2)$ and in the $B_{2g}$ $(xy)$ symmetry, respectively
(Fig.~\ref{stripes}). The LSCO system itself provides the
opportunity for a direct check. As found by neutron scattering,
the magnetic superstructure and, hence, the orientation of the
stripes rotates from diagonal to parallel when $x$ exceeds a
critical value of $x_s \simeq 0.055$  \cite{fuj02}.

In this Letter, we show that charged stripes can be directly
observed in a Raman experiment. The high intensity of the
additional response at low doping allows to determine the
temperature dependence of fluctuating stripes. The scaling
behavior raises implications on quantum criticality in the
cuprates.

A standard light scattering setup equipped with a scanning
spectrometer and the sample mounted on the cold finger of a
He-flow cryostat was used for the experiments. The scattering was
excited by the Ar$^+$ laser line at 458~nm. The absorbed power
ranged from 1 to 4~mW resulting in a local heating of 3 to 12~K as
determined by comparing energy gain and loss spectra. In the
figures, spot temperatures are indicated. The electronic Raman
response $\chi_{\mu}^{\prime \prime}(\omega,T) =
S_{\mu}(\omega,T)/\{1+n(\omega,T) \}$ is a two-particle
correlation function related to the conductivity
$\sigma^{\prime}(\omega,T)$. $\mu=B_{1g}$, $B_{2g}$ etc.,
$S_{\mu}(\omega,T)$, and $\{1+n(\omega,T) \}$ are the symmetry
index, the structure factor (proportional to the measured Raman
intensity), and the Bose factor, respectively. The symmetries
$\mu$ can be selected by the polarizations of the incident and the
scattered photons and correspond to form factors projecting out
different electron momenta. Hence, the response is a
momentum-sensitive transport quantity \cite{dev94,dev03}. In the
case of free carriers, the slope of the spectra at $\omega=0$,
$\tau_{0}^{\mu}(T)$, is proportional to a $\bf k$-resolved dc
conductivity.

\begin{figure}

{\centerline{\psfig{figure=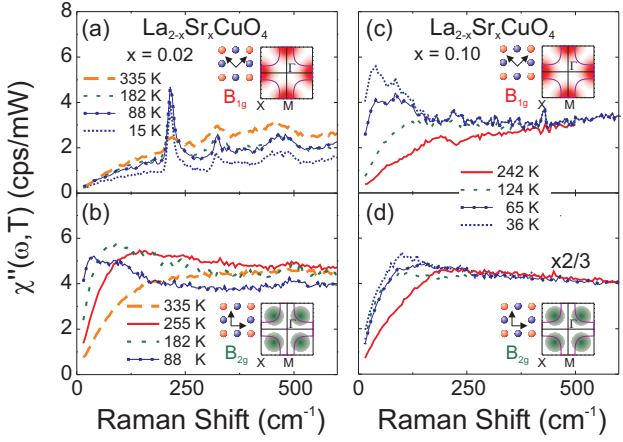,width=8.5cm}}}
\caption[]{Raman response $\chi_{\mu}^{\prime \prime}(\omega,T)$
of La$_{1.98}$Sr$_{0.02}$CuO$_4$ (a,b) and
La$_{1.90}$Sr$_{0.10}$CuO$_4$ (c,d). As indicated in the insets,
areas around the M points and the center of the quadrant are
projected out in $B_{1g}$ and $B_{2g}$ symmetry, respectively, on
a quadratic 2D lattice. The selection rules for 1D structures are
explained in Fig.~\ref{stripes}.} 

\label{data}
\end{figure}

\begin{figure}

{\centerline{\psfig{figure=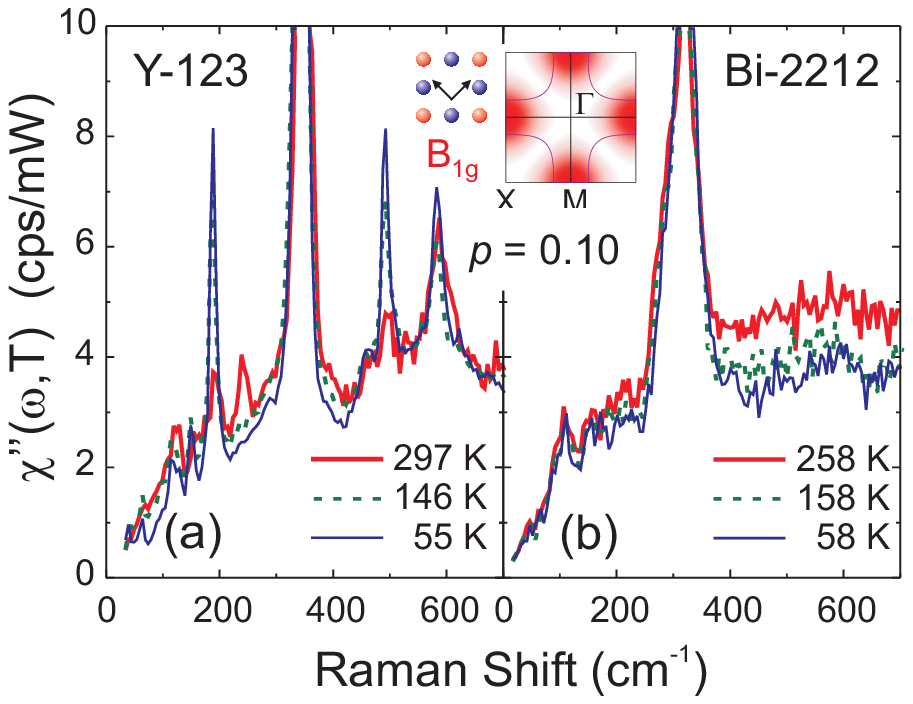,width=6.8cm}}}
\vspace{0.5cm} \caption[]{Raman response $\chi_{B_{1g}}^{\prime
\prime}(\omega,T)$ of  $\rm YBa_2Cu_3O_{6+x}$ (Y-123) and $\rm
Bi_2Sr_2CaCu_2O_{8+\delta}$ (Bi-2212) at $p=0.10$.}

\label{ybi}
\end{figure}

In Fig.~\ref{data}, the $B_{1g}$ and $B_{2g}$ spectra of LSCO at
$x=0.02$ and $x=0.10$ can be compared. At $x=0.10$ the $B_{2g}$
spectra [Fig.~\ref{data}~(d)] and their temperature dependence are
similar to those in $\rm Bi_2Sr_2CaCu_2O_{8+\delta}$ (Bi-2212) and
$\rm YBa_2Cu_3O_{6+x}$ (Y-123) \cite{ope00}. They exhibit an
increase of the slopes and hence a decrease of the dc scattering
rates (``Raman resistivities'') $\Gamma_{0}^{B_{2g}}(T)=
\hbar/\tau_{0}^{B_{2g}}(T)$ upon cooling. The results are
consistent with the optical and the dc resistivity to within 30\%
\cite{ope00}. For $B_{1g}$ symmetry [Fig.~\ref{data}~(c)] the
spectra are relatively flat at room temperature. Upon cooling a
dramatic increase of the initial slope occurs leading to a pile-up
of intensity at low energies quite similar to what is found in
ladder compounds \cite{goz03}. The $B_{1g}$ results in LSCO are in
clear contrast to those in Y-123 and Bi-2212 (Fig.~\ref{ybi})
which both show an insulating-type of behavior at $\omega
\rightarrow 0$ \cite{ope00,ven02a}. At low doping, $x=0.02$, the
$B_{1g}$ behavior [Fig.~\ref{data}~(a)] expected from
extrapolating the results of Y-123 and Bi-2212 is restored while
the initial slope of the $B_{2g}$ spectra [Fig.~\ref{data}~(b)]
increases as strongly as that of the $B_{1g}$ spectra at $x=0.10$
[Fig.~\ref{data}~(c)]. In either symmetry
[Fig.~\ref{data}~(a),~(b)] the overall intensity decreases upon
cooling indicating a reduction of the number of free carriers as
anticipated from the resistivity \cite{fuj99}. This is one of the
reasons why there is no pile-up of the $B_{2g}$ intensity at
$x=0.02$. Additionally, there is inhomogeneous broadening of the
peaks originating from both small variations of the Sr
concentration and local distortions due to the low temperature
structural transformation. At low temperature the stripes are at
least partially pinned which is equivalent to static order
\cite{fuj02a}.

According to the information from previous studies
\cite{ope00,ven02,ven02a} we have to conclude, that a new
scattering channel opens up at low temperature in the $B_{1g}$ and
the $B_{2g}$ symmetry at $x=0.10$ and $x=0.02$, respectively.
Since there is no sum rule as in the case of the optical
conductivity, the additional spectral weight is directly
superimposed on the response of the 2D ${\rm CuO_2}$ planes. At
$x=0.02$, the reduction of free carriers and the tendency to
static order reduces the effect.

\begin{figure}[b]
{\centerline{\psfig{figure=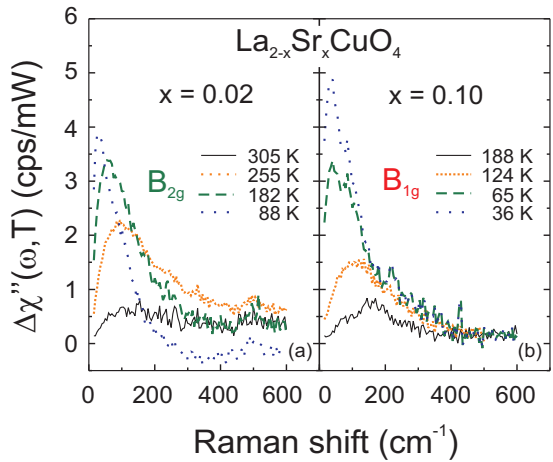,width=7.5cm}}}
\caption[]{Response of fluctuating charge order. A Drude-like peak
\cite{zaw90} with a characteristic energy $\Omega_c(x,T)$ is
revealed after subtraction of the 2D response of the ${\rm CuO_2}$
planes. At $x=0.02$ (a) and 0.10 (b) the additional response is
observed in $B_{2g}$ and $B_{1g}$ symmetry, respectively. The
styles of the lines (colours) do {\em not} correspond to similar
temperatures but rather highlight the {\em scaling} of the
response with temperature: similar spectra are obtained if the
temperatures differ by approximately a factor of 2.}

\label{stripe-response}
\end{figure}

The new peaks can be separated out to a good approximation by
subtracting the ``background'' of the 2D $\rm CuO_2$ planes. To
this end, we use analytic approximations to the high-temperature
spectra. In order to avoid any additional influence we assume that
the 2D response is independent of temperature and is determined by
the spectra at 335~K and 242~K for $x=0.02$ and $x=0.10$,
respectively. After subtraction, a Drude-like spectrum
\cite{zaw90} comparable to the one in ${\rm Sr_{14}Cu_{24}O_{41}}$
\cite{goz03} is obtained. The characteristic energy
$\Omega_c(x,T)$ corresponding to the peak position depends
strongly on temperature [Fig.~\ref{stripe-response}~(a),~(b)]. The
experimental peaks are narrower than expected for the Drude
response. In spite of the different scattering symmetries at
$x=0.02$ and 0.10 the spectral shapes are remarkably similar if a
factor of approximately 2 in the temperature scales is taken into
account [compare Fig.~\ref{stripe-response}~(a) and~(b)]. This
also indicates that the subtraction procedure has only little
influence.

\begin{figure}[b]
{\centerline{\psfig{figure=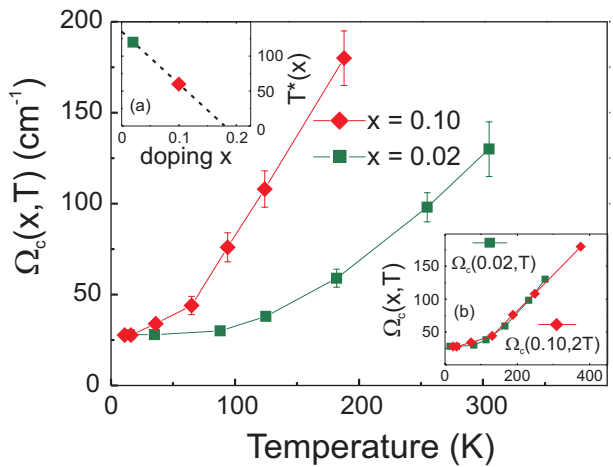,width=8cm}}}
\caption[]{Temperature dependence of the characteristic energy
$\Omega_c(x,T)$ of the stripe response (see
Fig.~\ref{stripe-response}) in LSCO. Inset (a) shows the
determination of the quantum critical point $x_c$ at which
$T^{\ast}(x)$ extrapolates to O (see text). From changing the
temperature scale (inset b) one obtains $T^{\ast}(x=0.02) \simeq
2T^{\ast}(x=0.10)$. Inset (b) demonstrates that temperature is the
{\it only} scale of $\Omega_c(x,T)$. For the two doping levels
studied the scaling factor is approximately 2.} \label{QCP}
\end{figure}

In Fig.~\ref{QCP}, $\Omega_c(x,T)$ is plotted for the two samples
studied. $\Omega_c(x,T)$ saturates below crossover temperatures
$T^{\ast}(x)$ of approximately 120 and 60~K for $x=0.02$ and 0.10,
respectively, at a doping-independent energy of 25--30~${\rm
cm^{-1}}$.

The results from the two doping levels can be mapped on top of
each other by scaling the temperature by a factor of approximately
2, $\Omega_c(0.1,2T) \simeq \Omega_c(0.02,T)$ (inset (b) of
Fig.~\ref{QCP}). This implies and , hence,  corroborates
$T^{\ast}(0.02) \simeq 2T^{\ast}(0.10)$ as derived from
$\Omega_c(x,T)$ directly (see inset (a) of Fig.~\ref{QCP}). This
is a strong indication that, independent of symmetry, the observed
features originate from fluctuation phenomena where temperature is
the only energy scale and doping is the control parameter
\cite{var89,sac00,and01}. For a quantum phase transition
$T^{\ast}(x)$ should approach 0 at the quantum critical point
(QCP) $x_c$, $T^{\ast}(x \rightarrow x_c)=0$. For $x < x_c$ the
cross-over temperature $T^{\ast}(x)$ separates the fluctuation
regime at $T>T^{\ast}$ from an at least partially ordered state at
$T<T^{\ast}$. In the case of long range static order such as in an
antiferromagnetic N\'eel state the intensity of the new
fluctuation mode is expected to vanish below $T^{\ast}(x)$.

In the presence of a QCP, $\Omega_c(x,T)$ is a quantity
proportional to the mass $M(x,T)$ in the fluctuation propagator
\cite{and01,dic04,cap05}. At sufficiently high temperatures, where
thermal and quantum fluctuations dominate, the mass $M(x,T)$ and,
consequently, $\Omega_c(x,T)$ are linear in $T$, $\Omega_c(x,T)
\propto M(x,T)=\alpha (x) [T-T^{\ast}(x)]$. Below $T^{\ast}(x)$ in
the (partially) ordered state, $M(x,T<T^{\ast})$ saturates
\cite{and01}. Hence, the extrapolation to zero of the linear part
of $\Omega_c(x,T)$ or $M(x,T)$ should provide estimates of
$T^{\ast}(x)$. In fact, the data at $x=0.10$ have a linear part
above approximately 120~K (Fig.~\ref{QCP}). At $x=0.02$, the
linear part of $T^{\ast}(x)$ is not as evident since the cross
over is probably closer to the highest measuring temperature than
at $x=0.10$. Therefore, in order to obtain an estimate of $x_c$ we
use $T^{\ast}(0.02) \simeq 2T^{\ast}(0.10)$ (see above). With the
relation $T^{\ast}(x) \propto x$ as supported by both theory and
experiment \cite{and01,sta99,tallon00,Gutmann} one arrives at
$x_c=0.18 \pm 0.02$ (inset (a) of Fig.~\ref{QCP}). This analysis
must be qualified by noting that the symmetry of the two
``stripe'' states is different, and thus the two data points in
panel (a) might be related in a more complicated way.

It is tempting to identify $T^{\ast}(x)$ with the pseudogap line.
Apparently, it is close to the lower boundary of values found
experimentally \cite{sta99,Gutmann}. If we assume that
fluctuations dominate the physics, then $T^{\ast}(x)$ is defined
as the temperature at which the mass $M$ extrapolates to zero. As
a consequence, $T^{\ast}(x)$ depends on the characteristic cut-off
of the respective experimental probe and cannot be determined
uniquely \cite{and01}.

The results in LSCO provide direct experimental evidence of local
symmetry breaking  by the formation of quasi-1D structures (cf.
Fig.~\ref{stripes}) at low temperature and doping which can be
interpreted in terms of a charge ordering instability. The main
support comes from the new type of response [see
Fig.~\ref{stripe-response}~(a),~(b)] superimposed on the usual
spectra of the $\rm CuO_2$ planes and its dependence on symmetry
demonstrating the reorientation of the charge modulation as a
function of doping as already suggested by neutron scattering on
the spin system \cite{fuj02}. At either doping the new response is
unrelated to the dc resistivity indicating its dynamical and most
probably transverse nature.

It is an important question as to why the low-energy response
cannot be observed in Y-123 and Bi-2212 at $p=0.10$. We
hypothesize that the correlation length $\xi_s$ of the ordering
phenomenon must exceed a minimal value to make the response
visible. For example in the case of phonons, translational
symmetry must be established over several lattice constants to
guarantee the selection rules to hold. Here, this seems to be a
lower bound, and, more likely, the electronic mean free path
$\ell$ is the relevant scale. On the other hand, $\xi_s$
corresponds to a fluctuation frequency of order $\Omega_c \propto
(\xi_s)^{-z}$ ($z=2$ for damped modes). If $\xi_s$ is
substantially smaller than in LSCO $\Omega_c$ is expected to be
larger. As a consequence, the intensity becomes weaker and is
distributed over a bigger energy range. Then, the response of the
stripes cannot be separated from the usual one of the planes.

We conclude that the charge-ordering instability may be an
intrinsic feature of the copper-oxygen plane. The maximal doping
level up to which the stripes can be observed directly depends on
whether or not the lattice helps to stabilize the order. In LSCO
$\xi_s$ has obviously the proper magnitude to allow the
observation of the stripes in an optical experiment facilitating a
detailed study of the dynamical and critical behavior. If part of
the La is substituted by Nd or Eu \cite{tra95,kla00} or if Sr is
replaced by Ba  \cite{fuj02a} $\xi_s$ increases, and static order
is established by a modification of the tilt of the ${\rm CuO_6}$
octahedra or, equivalently, by slightly changing the corrugation
of the copper-oxygen plane. Above a critical tilt angle
superconductivity is quenched \cite{kla00}. Apparently, if $\xi_s$
increases the superconducting transition temperature $T_c$
decreases suggesting a relation between charge ordering and
superconductivity. Since the lattice determines the correlation
length $\xi_s$ it influences $T_c$, but, according to the results
here, in a subtle and indirect way.

We would like to express our gratitude to L. Benfatto, C.
Di~Castro, T.P. Devereaux and M. Grilli for important discussions
and Ch. Hartinger for critically reading the manuscript. The
project has been supported by the DFG under grant-no. Ha2071/2-2.
F.V. and Q.-M.Z. would like to thank the Gottlieb Daimler - Karl
Benz Foundation and the Alexander von Humboldt Foundation,
respectively.

\end{document}